\documentclass[sigconf]{acmart}

\AtBeginDocument{%
  }

\acmConference[LOCO 2024]{1st International Workshop on Low Carbon Computing}{December 03, 2024}{Glasgow, UK/Online}
\acmISBN{}
\acmDOI{}
\setcopyright{none}

\usepackage{multirow}
\begin{document}

\title{Carbon-Aware Microservice Deployment for Optimal User Experience on a Budget}


\author{Kevin Kreutz}
\authornotemark[1]
\email{kreutz@campus.tu-berlin.de}
\affiliation{%
  \institution{Technical University of Berlin}
  \city{Berlin}
  \country{Germany}
}

\author{Philipp Wiesner}
\authornote{All authors contributed equally to this research.}
\email{wiesner@tu-berlin.de}
\orcid{xxxxxxxx}
\affiliation{%
  \institution{Technical University of Berlin}
  \city{Berlin}
  \country{Germany}
}

\author{Monica Vitali}
\authornotemark[1]
\email{monica.vitali@polimi.it}
\orcid{0000-0002-5258-1893}
\affiliation{%
  \institution{Politecnico di Milano}
  \city{Milan}
  \country{Italy}
}


\begin{abstract}
The carbon footprint of data centres has recently become a critical concern. 
So far, most carbon-aware strategies have focused on leveraging the flexibility of scheduling decisions for batch processing by shifting the time and location of workload executions. 
However, such approaches cannot be applied to service-oriented cloud applications, since they have to be reachable at every point in time and often at low latencies.

We propose a carbon-aware approach for operating microservices under hourly carbon budgets.
By choosing the most appropriate version and horizontal scaleout for each microservice, our strategy maximizes user experience and revenue while staying within budget constraints.
Experiments across various application configurations and carbon budgets demonstrate that the approach adapts properly to changing workloads and carbon intensities.
\end{abstract}





\maketitle

\section{Introduction}\label{sec:intro}
Governing bodies like the European Union (EU) are activating several actions aiming at reaching carbon neutrality by 2050~\cite{fetting2020european}. The ICT sector plays an important role in this challenging task, contributing to 3.9\% of the overall global emissions with an increasing trend over the years~\cite{ICTStandard}. To this aim, the European Green Digital Coalition (EGDC) has involved several companies in the joint effort to measure and reduce the environmental impact of ICT, with a target for carbon neutrality by 2040~\cite{declaration2021green}. Similarly, the EU Emissions Trading System (EU ETS)~\cite{ets} promotes carbon neutrality targets by capping the total amount of emissions allowed for specific sectors. 

Cloud providers have announced carbon neutrality goals~\cite{amazonRep, googleRep, azureRep}. Although energy efficiency has improved at the infrastructure level, cloud emissions continue to rise due to the increasing demand for computational power from hosted applications. It is therefore important to reduce the environmental impact of these applications. To achieve this, enhancing applications with adaptive features~\cite{meza2023defcon, vitali2022towards} can help reduce their emissions when needed.

In the proposed approach, we apply the concept of hourly budgets of carbon emissions or time-based renewable energy certificates~\cite{TEACS, Hourly_EAC} to microservice architectures.
We propose an algorithm able to continuously reconfigure the microservice deployment according to the current budget, the energy mix of the electricity grid, and the workload. 
We show that our approach can satisfy the budget constraints while maximizing the user experience and the revenue for the application owner. 


\section{Approach}\label{sec:approach}

The proposed approach includes carbon-awareness in the deployment process of long-running cloud applications, leveraging flexibility~\cite{6128960, 10.1145/3464298.3493399,sukprasert2023quantifying,10.1007/s11227-023-05506-7, 9860626, wiesner2024qualitytime} to adjust application deployment and scaling based on the current carbon intensity of the electricity grid.

\begin{figure}[t]
    \centering
        \includegraphics[width=1\columnwidth]{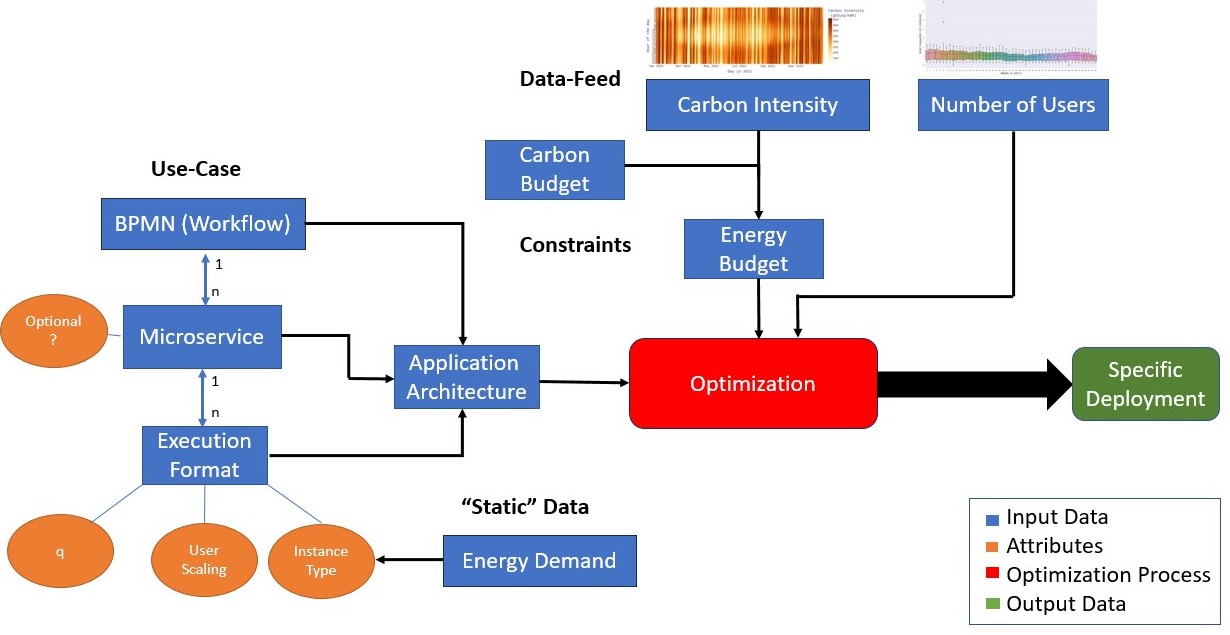}
        \caption{Overview of the Approach}
    \label{fig:approach-overview}
    \Description[Overview of the approach]{The schema of the approach. The core is an optimization which outputs an optimal deployment for an application. It takes as input the application description, the energy budget, and the workload expressed as the number of users expected in the period.}
\end{figure}

An overview of the approach is shown in~\autoref{fig:approach-overview}. At its core is an optimisation algorithm, 
which selects the best configuration of an application to avoid exceeding the carbon budget while optimising other non-functional requirements, such as the Quality of the Experience (QoE) of the users and the revenue of the application owner. In this section, we provide additional details of the components of the approach and we model the objective function used for the optimisation.

\paragraph{Application Architecture} The type of applications considered in this work are long-running cloud-native applications composed of multiple microservices. The application workflow can be modelled and enriched using BPMN as discussed in \cite{vitali2022towards}. Each microservice can be provided in multiple versions, each one with different characteristics in terms of energy demand and quality of experience provided to the users. For instance, a low-power version of a microservice might remove some computationally demanding functionalities affecting the overall satisfaction of the user. A microservice can also be annotated as optional. Optional microservices are not necessary to reach the overall goal of the application. Still, they can increase the QoE of the user and/or generate rewards for the application owner (e.g., advertisement). 

\paragraph{Energy Budget} It expresses the carbon-related information including the carbon budget assigned to the application and the carbon intensity of the energy grid. We assume the carbon budget is assigned yearly but it is dynamically split hourly according to the expected workload. The carbon intensity is the current mix of green and brown energy in the electricity grid. This value can be used to compute how much energy can be used in the time slot without exceeding the carbon budget.

\paragraph{Number of Users} The application can be associated with an estimated number of requests (workload) during the year according to the information collected in the previous years.

The objective function can be expressed as in Equation~\ref{eq:optFun}
\begin{equation}\label{eq:optFun}
    \max \left(\alpha \sum_{ms \in CA}{\frac{QoE_{ms}}{\#ms}} + \beta \sum_{ms \in CA}{\frac{rev_{ms}}       {revMax}}\right),
\end{equation}
where $QoE_{ms}$ is the QoE associated with a microservice version, $rev_{ms}$ is the revenue for the application owner associated with a microservice, $\alpha$ and $\beta$ are two weights associated to the relevance of these to aspects for the application owner. The complete implementation of the optimization algorithm is available to the public\footnote{\url{https://github.com/POLIMIGreenISE/carbonbudget.git}}.

\section{Experiments and Results}\label{sec:results}
\begin{figure}[t]
    \centering       
    \includegraphics[width=0.95\columnwidth]{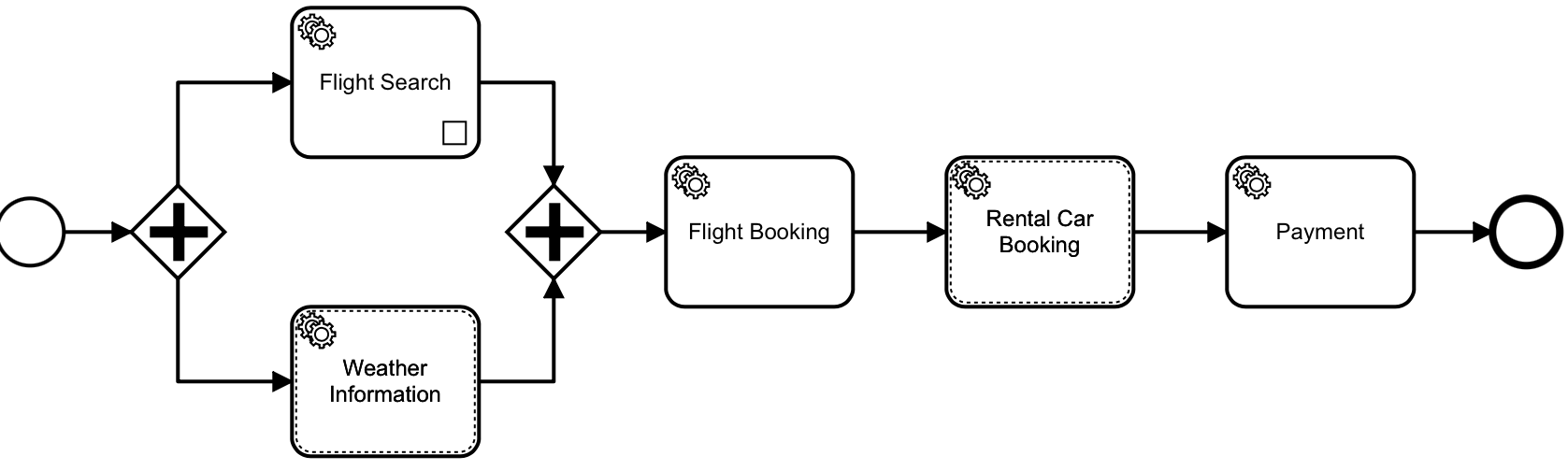}
    \caption{Flight Booking application in BPMN}
    \Description[Workflow of a flight booking application]{The flight booking application is composed of five microservices. Some are annotated as optional and others have multiple versions.}
    \label{fig:flight-booking}
\end{figure} 

\begin{table}[t]
    \centering
    \resizebox{\columnwidth}{!}{%
    \begin{tabular}{|c|c|c|c|c|c|c|c|c|}
         \hline
         \textbf{Microservice} & \textbf{Versions}& \textbf{Optional} &  \textbf{Instance-Type} & \textbf{ED} & \textbf{q} & \textbf{uc} & \textbf{QoE} & \textbf{rev}\\
         \hline\hline
         \multirow{3}{*}{Flight Search}& Low Power & \multirow{3}{*}{no} & t3.micro & 13.0 & 0.5 & 20000 or 5000& 0.5 or 0.1& 0\\
         & Normal  & & t3.xlarge & 39.9 & 0.7 & 20000 or 5000& 0.75 or 0.3& 0\\
         & High Performance & &  g2.2xlarge &305.4 & 0.9 & 20000 or 5000& 1 & 0\\
         \hline
         
         \multirow{2}{*}{Weather Information}& Off & \multirow{2}{*}{yes} &  & 0.0 & 0.9 &  & 0 & 0\\
         & Normal  & & t3.micro & 13.0 & 1.0 & 20000 or 5000& 1 & 0.2\\
         \hline
         
         \multirow{2}{*}{Flight Booking}& Low Power & \multirow{2}{*}{no} & t3.micro & 13.0 & 0.5 & 20000 or 5000& 0.5 or 0.1& 0\\
         & Normal  & & t3.xlarge & 39.9 & 0.95 & 20000 or 5000& 1 & 0\\
         \hline
         
         \multirow{3}{*}{Rental Car Booking}& Off & \multirow{3}{*}{yes} &  & 0.0 & 0.75 &  & 0 & 0\\
         & Normal  & & t3.xlarge & 39.9 & 0.9 & 20000 or 5000& 0.75 or 0.3& 2\\
         & High Performance & &  g2.2xlarge & 305.4 & 1.0 & 20000 or 5000& 1 & 2\\
         \hline
         
         Payment & Normal & no & t3.xlarge & 39.9 & 0.99 & 20000 or 5000& 1 & 0\\
         \hline
    \end{tabular}
    }
    \caption{Flight Booking microsevices' annotations}
    \label{tab:parameters}
\end{table}

We validated the approach with a simple application, shown in \autoref{fig:flight-booking} using an enriched BPMN~\cite{vitali2023enriching}. The application consists of five microservices. Two of them (\textit{Weather Information} and \textit{Rental Car Booking}) are optional. The \textit{Flight Search} microservice is instead provided with multiple versions. The specific annotations of the five microservices are summarised in~\autoref{tab:parameters}, including versions and the optional label. For each version, the required instance type for deployment is specified along with the estimated energy demand ($ED$). Each version is linked to a revenue ($rev$, applicable only to optional functionalities) and a $QoE$ value. The parameter $uc$ represents the number of user requests handled by a single instance, while $q$ denotes the percentage of users who move on to the next microservice. In the experiments, most parameters were kept fixed, while varying the values of $uc$ and $QoE$.

The proposed algorithm (\textit{Optimal Selection (OS)}), is designed to choose the best configuration for each microservice, ensuring budget constraints are met while optimizing the objective function (\autoref{eq:optFun}). It is compared against three baselines: (i) \textit{High Performance (HP)} - all microservices are deployed in high-performance versions and optional microservices are deployed by default; (ii) \textit{Sequential Carbon-Aware (SCA)} - the application adjusts based on the available budget using three fixed sequential configurations for all microservices; and (iii) \textit{Simple Carbon-Aware (CA)} - similar to \textit{SCA}, but the three configurations do not have to be sequential.

\begin{figure}[t]
    \centering
    \includegraphics[width=1\columnwidth]{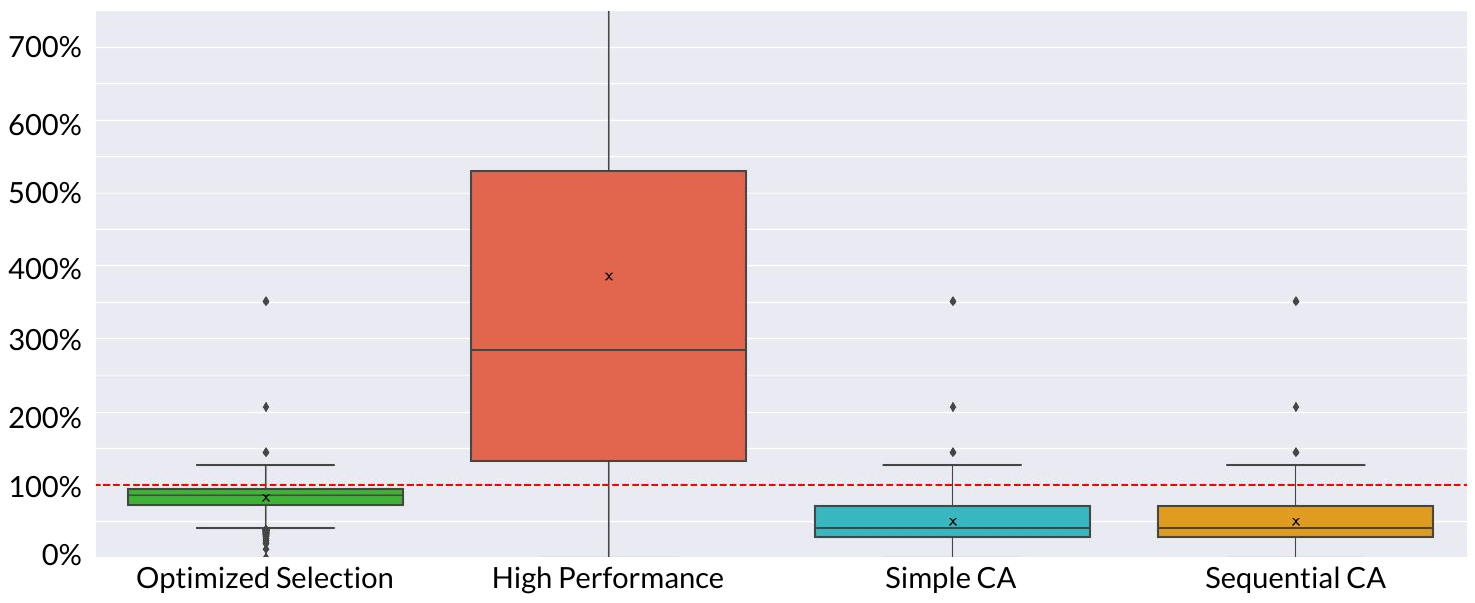}
    \caption{Carbon budget utilisation}
    \label{fig:averageBudget}
    \Description[Comparison between the proposed approach and the baselines]{The picture shows the box plot of the average carbon budget utilisation of the different baselines compared with the proposed solution showing a better capacity of the proposed solution to fulfil the budget constraints while maximizing its utilisation.}
\end{figure}

Experiments were conducted using different carbon budgets. For instance, \autoref{fig:averageBudget} shows the results obtained with a carbon budget set between the emissions needed to operate the application in high-performance mode and those required for low-power mode over the course of a year. As can be observed, the $HP$ baseline significantly violates the carbon budget, while all the other solutions can fulfil the budget constraints. The proposed approach effectively maximizes budget utilization, resulting in higher revenue and QoE. Compared to the carbon-aware baselines, the OS algorithm delivers an average QoE improvement of 5.44\% and a revenue increase of 32.38\%.

\section{Conclusion and Future Work}\label{sec:conclusion}

We proposed a carbon-aware approach to adapt the operation of cloud-native applications based on the availability of renewable energy, a carbon budget, and the current workload, while maximizing both QoE and revenue. Our approach was tested under dynamic workloads and fluctuating energy mixes, demonstrating its capability to meet the carbon budget while maximizing its utilization, leading to higher QoE and revenue. In future work, we plan to explore additional configurations and experiment with different weights for revenue and QoE in the objective function.

\begin{acks}
This work was supported by the project FREEDA, funded by the frameworks PRIN (MUR, Italy) and Next Generation EU (CUP: I53D23003550006) and by the Spoke 1 ``FutureHPC \& BigData'' of the Italian Research Center on High-Performance Computing, Big Data and Quantum Computing (ICSC) funded by MUR Missione 4 - Next Generation EU (NGEU).
\end{acks}


\bibliographystyle{ACM-Reference-Format}
\bibliography{references}


\end{document}